\documentclass[aps,prd,reprint,superscriptaddress,longbibliography]{revtex4-2}

\usepackage[colorlinks=true,linkcolor=blue,citecolor=blue,urlcolor=blue]{hyperref}
\usepackage{microtype} 
\usepackage{graphicx} 
\usepackage{amsfonts, amsthm, amsmath, amssymb, physics, upgreek, stackrel} 
\usepackage{array, enumitem} 
\usepackage{verbatim, listings} 
\usepackage[x11names,dvipsnames]{xcolor}
\usepackage{dsfont, xr, color,bm} 
\usepackage{placeins} 
\usepackage{orcidlink}

\usepackage{JanShortcuts}
\newcommand{\orcidLouw}{\orcidlink{0000-0002-5111-840X}}

\definecolor{Red}{RGB}{232,21,16}

\definecolor{StrangeGreen}{HTML}{58B359}
\def\Jj{\ensuremath{{\mathcal{J}}}}
\definecolor{Red}{RGB}{232,21,16}
\definecolor{StrangeGreen}{HTML}{58B359}

\begin{document}

	\title{Analytical solutions to the non-equilibrium Green's functions in large-$q$ SYK models}
	
	\author{Jan C. Louw \orcidLouw}
	\affiliation{Arbeitsgruppe Computing (AG C), Gesellschaft f\"ur wissenschaftliche Datenverarbeitung mbH G\"ottingen (GWDG), Burckhardtweg 4, D-37077 G\"ottingen, Germany}
	\email{jancillie.louw@gwdg.de}
	
	\begin{abstract}
		It is known that the Sachdev-Ye-Kitaev (SYK) model is exactly solvable at leading order in $1/q$; specifically, the instantaneously thermal time block after the quench is known in closed form. Thus far the remaining (inherently out of equilibrium) blocks have only been numerically accessible. Here we provide their analytical solutions for a quench between non-commuting $q/2$-body SYK Hamiltonians. We obtain the Green's functions in closed form for every time block. We extract a simple relation between the pre- and post-quench energies $\epsilon_1 \propto \epsilon_0$ leading to an exact temperature update rule. Via a Cauchy-Schwarz inequality, we show that heating is inevitable for the closed system; hence, the second law emerges geometrically. The solution is used in a companion paper to address what ``instantaneous thermalization'' means at finite $q$ \cite{CompanionOsterkorn}.
	\end{abstract}
	
	\maketitle
	
    \section{Introduction}
	The Sachdev-Ye-Kitaev (SYK) model is a rare example of a strongly interacting quantum system that is maximally chaotic yet still solvable \cite{SachdevYe1992,Chowdhury2022Sep,Maldacena2016Aug, Maldacena2016Nov, Eberlein:2017jb,Louw2022Feb, Kitaev2015}. The solvability is in the low-temperature conformal limit and at leading order in $1/q$ (at all temperatures) \cite{Maldacena2016Nov, Eberlein:2017jb,Louw2022Feb}. The conformal nature also includes a holographic dual with near-extremal black holes \cite{Kitaev2015,Sachdev2015Nov,Louw2023Feb,Louw2023Oct}. This combination makes it the perfect model for questions about thermalization that are otherwise inaccessible in the  macroscopic limit \cite{Almheiri2024Aug,Louw2022Feb}. Nonequilibrium protocols within and between SYK models have accordingly been studied in a range of settings \cite{Sohal2022Jan, Bhattacharya2019Jul,Haldar2020Feb,Larzul2022Jan,Bandyopadhyay2023May,Eberlein:2017jb,Louw2022Feb,Kuhlenkamp2020Mar,Jaramillo2025May,Perugu2025Nov}.
	
	The large-$q$ solvability is almost in a class of its own when it comes to thermalization. In the diagonal blocks, both operators evolve under the same Hamiltonian. In this block the solutions reduce to a thermal form; thus, it is said to thermalize instantaneously. The off-diagonal blocks, with the times $t_1,t_2$ being pre- and post-quench, respectively, are however inherently non-thermal since they involve evolution under two distinct Hamiltonians. Existing closed-form results stop short of these blocks. In prior studies \cite{Eberlein:2017jb,Louw2022Feb}, a kinetic term is present before the quench, and only the post-quench block is accessible analytically; the off-diagonal parts are computed numerically. 
    
    Here, we close this gap for a quench between non-commuting $q/2$-body SYK Hamiltonians. This non-trivial setup turns out to be fully solvable in all blocks. We find that the off-diagonal blocks encode the memory of the quench and so the exact relation between pre- and post-quench temperatures. From the geometry of the coupling, a Cauchy-Schwarz inequality bounds the temperature update so that the system can only heat up. These solutions enable quantitative comparisons with finite-$q$ numerics in the quench region. As shown in a companion paper \cite{CompanionOsterkorn}, the off-diagonal block reconciles the instantaneous thermalization in the large-$q$ limit with the finite thermalization rate observed in numerics.
	
	\section{Model} \label{secModel}
	
	We consider a single SYK dot of $N$ spinless complex fermions subject to two mutually non-commuting $q/2$-body interactions with time-dependent strengths,
	\begin{equation}
		H(t)
		= \sum_{i} \hspace{-1mm} \sum\limits_{\substack{ \{\bm{\mu}\}_1^{q/2} \\ \{\bm{\nu}\}_1^{q/2} }} \hspace{-2mm}Z^{(i)}(t)^{\bm{\mu}}_{\bm{\nu}} \hat{c}^{\dag}_{\mu_1} \cdots \hat{c}^{\dag}_{ \mu_{q/2}} \hat{c}_{\nu_{q/2}}^{\vphantom{\dag}} \cdots \hat{c}_{\nu_1}^{\vphantom{\dag}},
		\label{hi}
	\end{equation}
	where $\{\bm{\nu}\}_{1}^{q/2} \equiv 1\le \nu_1<\cdots< \nu_{q/2}\le\Nn$, and $\hat{c}^{\dag}_{\alpha}, \hat{c}_{\alpha}$ create and annihilate a fermion of flavor $\alpha$. The couplings $Z^{(i)}(t)^{\bm{\mu}}_{\bm{\nu}}$ are independent Gaussian random variables with zero mean and variance $\overline{|Z^{(i)}|^2} = (N/2)^{1-q}[(q/2)!\Jj^{(i)}/q]^2$, drawn independently for each $i$, $\bm{\mu}$ and $\bm{\nu}$. In particular one writes out the random variables as
	\begin{equation}
		Z^{\bm{\mu}}_{\bm{\nu}} = \cos(\theta/2) X^{\bm{\mu}}_{\bm{\nu}} + \imath \sin(\theta/2) Y^{\bm{\mu}}_{\bm{\nu}},
		\label{Zdecomp}
	\end{equation}
	where $X$ and $Y$ are real and of equal variance. The disorder average of the action receives contributions from both $\overline{|Z^{\bm{\mu}}_{\bm{\nu}}|^{2}}$ and $\overline{Z^{\bm{\mu}}_{\bm{\nu}} Z^{\bm{\nu}}_{\bm{\mu}}}$, and the anomalous combination
	$\overline{(Z^{\bm{\mu}}_{\bm{\nu}})^{2}} = \cos\theta \; \overline{(X^{\bm{\mu}}_{\bm{\nu}})^{2}}$ signals a superconducting instability \cite{chowdhury_berg_2020}. Setting $\theta = \pi/2$ removes it. What remains is the relation between $Z^{\bm{\mu}}_{\bm{\nu}}$ and $Z^{\bm{\nu}}_{\bm{\mu}}$, which we may parametrize by a phase, $Z^{\bm{\nu}}_{\bm{\mu}} = e^{\imath \phi} (Z^{\bm{\mu}}_{\bm{\nu}})^{*}.$
	We consider the Hermitian case ($\phi = 0$) which is reflected in real couplings $\vec{\Jj}$.
	
\begin{figure}[ht]
		\centering
		\begin{minipage}[b]{0.54\columnwidth}\centering
			\begin{tikzpicture}[>=latex]
    \def\Ta{1.5}         
    \def\Tw{2}         
    \def\ya{0.0}         
    \def\yb{1.1}         
    \fill[orange!30] (0,-0.35) rectangle (\Tw,1.75);
    \draw[->,thick] (-\Ta,0) -- (\Tw,0) node[right] {$t$};
    \draw[->,thick] (-\Ta,-0.35) -- (-\Ta,1.75)
        node[above,align=center] {$\vec{\mathcal{J}}(t)$};
    \draw[dashed,gray!70] (0,-0.35) -- (0,1.75);
    \node[below] at (0,-0.35) {$\tau_1 = 0$};
    \draw[very thick,RoyalBlue3] (-\Ta,\ya) -- (0,\ya);
    \draw[very thick,RoyalBlue3,dotted] (0,\ya) -- (0,\yb);
    \draw[very thick,RoyalBlue3] (0,\yb) -- (\Tw,\yb);
    \filldraw[RoyalBlue3] (0,\yb) circle (1.4pt);
    \draw[RoyalBlue3,fill=white] (0,\ya) circle (1.4pt);
    \node[left] at (-\Ta,\ya) {$\vec{\mathcal{J}}_{0}$};
    \node[left] at (-\Ta,\yb) {$\vec{\mathcal{J}}_{1}$};
    \node[RoyalBlue3] at (-0.5*\Ta,\ya+0.32) {$H_0$};
    \node[RoyalBlue3] at (0.5*\Tw,\yb+0.32) {$H_1$};
    \node[gray!70] at (-0.5*\Ta,-0.75) {$I_0$};
    \node[gray!70] at (0.5*\Tw,-0.75) {$I_1$};
    \node[anchor=south west,inner sep=1pt] at (0.03,\ya+0.45)
        {\scriptsize $c_{01} \propto \vec{\mathcal{J}}_0\cdot\vec{\mathcal{J}}_1$};
\end{tikzpicture}\\[2pt](a)
		\end{minipage}\hfill
		\begin{minipage}[b]{0.44\columnwidth}\centering
			\begin{tikzpicture}[scale=1]
    \def\T{1}
    \def\R{1}
    \draw[->, thick] (-1*\T,-1*\T) -- ({(\R+0.8)*\T},-1*\T) node[right] {$t_1$};
    \draw[->, thick] (-1*\T,-1*\T) -- (-1*\T,1.8*\T) node[above] {$t_2$};

    \fill[orange!15] (0*\T,-1*\T) rectangle (1.3*\T,0*\T);  
    \fill[orange!15] (-1*\T,0*\T) rectangle (0*\T,1.4*\T);  
    \fill[orange!30] (0*\T,0*\T) rectangle (1.3*\T,1.4*\T); 

    \draw[black!30, thick, dashed] (-1*\T,-1*\T) -- (1*\T,1*\T);
    
    \draw[RoyalBlue3, line width=1pt] (0*\T,-1*\T) -- (0*\T,1.4*\T);
    \draw[RoyalBlue3, line width=1pt] (-1*\T,0*\T) -- (1.3*\T,0*\T);
    
    \draw[black, line width=1pt, dashed] (1.3*\T,-1*\T) -- (1.3*\T,1.4*\T);
    \draw[black, line width=1pt, dashed] (-1*\T,1.4*\T) -- (1.3*\T,1.4*\T);
    
    \node[fill=white, fill opacity=0.8, text opacity=1, inner sep=2pt] at (-0.5*\T,-0.5*\T) {$(0,0)$};
    \node[fill=white, fill opacity=0.8, text opacity=1, inner sep=2pt] at (0.65*\T,-0.5*\T) {$(1,0)$};
    \node[fill=white, fill opacity=0.8, text opacity=1, inner sep=2pt] at (-0.5*\T,0.7*\T) {$(0,1)$};
    \node[fill=white, fill opacity=0.8, text opacity=1, inner sep=2pt] at (0.65*\T,0.7*\T) {$(1,1)$};
    
    \node[below] at (0,-1*\T) {$\tau_1 = 0$};
    \node[left] at (-1*\T,0) {$\tau_1$};
\end{tikzpicture}\\[2pt](b)
		\end{minipage}
		\caption{(a) Schematic of the quench protocol. The time axis is partitioned into $I_0$, $I_1$. The quench occurs at $\tau_1 = 0$. 
			The quench strengths are encoded in the alignment of the coupling vectors, $c_{ab} \equiv \vec{\mathcal{J}}_a\cdot\vec{\mathcal{J}}_b/(\mathcal{J}_a \mathcal{J}_b)$. To simplify notation, we label functions by the intervals in which their time arguments lie, e.g. $\Jj(t_1)=\Jj_a$ or $c_{ab} = c(t_1,t_2)$ for $(t_1,t_2)\in I_{b}\! \times\! I_{b}$.
			(b) Partition of the $(t_1,t_2)$-plane into blocks $I_{b} \times I_{b}$. The boundary conditions are imposed along the blue connecting lines. 
		}
		\label{fig:protocol}
	\end{figure}
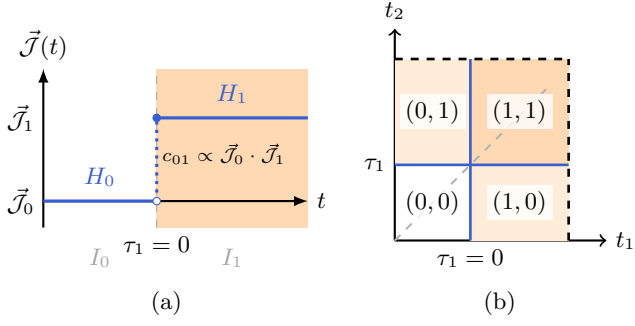

	The protocol, depicted in Fig.~\ref{fig:protocol}.(a), is parametrized by a vector of effective coupling constants $\vec{\Jj}(t) = (\Jj^{(0)}(t), \Jj^{(1)}(t),\ldots)$ corresponding to each individual SYK Hamiltonian. We denote the norm as $\Jj(t) = \vert\vert\vec{\Jj}(t)\vert\vert$. Throughout this work, a superscript on $\Jj^{(i)}$ labels a component of this vector, while a subscript labels the time interval in which the argument lies.
	
	The essential feature of \eqref{hi} is that the individual terms do not commute. Consequently, the disorder-averaged kernel $\vec{\Jj}(t_1)\cdot \vec{\Jj}(t_2)$ does not factorize into a product of a function of $t_1$ and a function of $t_2$. Had we instead taken a single SYK term with a time-dependent coupling, $\vec{\Jj}(t_1)\cdot \vec{\Jj}(t_2) $ would factorize, and the ``quench'' would amount to nothing more than a time-dependent choice of energy units, leaving the system in equilibrium throughout. It is the non-factorizing part of $\vec{\Jj}(t_1)\cdot \vec{\Jj}(t_2) $ that makes the problem genuinely out of equilibrium, and it is exactly this object that controls both the off-diagonal correlations, see Fig.~\ref{fig:protocol}.(b), and the amount of heating.
	
	\section{Quench dynamics} \label{DerivationKB}
	We consider time evolution along the Keldysh contour $\Cc$, with focus on the flavor-averaged correlation functions $\Gg(t_1,t_2) \equiv -\imath\ex{\Tt_{\Cc} \hat{c}(t_1) \hat{c}^{\dag}(t_2)}$. In the large $q$ limit they are captured by the expressions
	\begin{equation}
		\Gg(t_1,t_2)  = - \imath\, e^{ g(t_1,t_2)/q}/2, \label{largeqform}
	\end{equation}
    for $t_1>_{\Cc} t_2$, meaning that the interactions enter in at first order in $1/q$. For a longer discussion on the above see App.\ref{AppGreen}. Although we are considering the charged/complex SYK case \cite{Sachdev2015Nov,Davison2017Apr,Gu2020Feb}, we will focus on half filling, which overlaps with the Majorana SYK case in many ways. For instance $\Gg(t_2,t_1) = -\Gg(t_1,t_2)^*$ in our charge-neutral case \cite{Eberlein:2017jb,Louw2022Feb}.  At equal times we have the boundary condition $g(t,t) = 0$. This boundary condition stems from the fact that at equal times the Green's functions are expectation values of the form $\langle \hat{A}(t)\hat{A}^\dag(t)\rangle = \langle (\hat{A}\hat{A}^\dag)(t)\rangle$. Then, if the system is thermal or $\hat{A}\hat{A}^\dag$ is conserved, we will have time independence along the diagonal time. For complex fermions the reason for this is that $\hat{A} = \hat{c}$ and the above summed over $i$ reduces to the total charge density which is conserved. For Majorana fermions, $\hat{A} = \hat{\chi}$, which squares to a scalar, thus trivially conserved.
	
	Via the known SYK effective action, one obtains the self-energy in the thermodynamic limit $N\to\infty$
	\begin{align}
		q\Upsigma(t_1,t_2)
		=& -2\vec{\Jj}(t_1) \cdot \vec{\Jj}(t_2)\, \notag\\ &\cdot [4\Gg(t_1,t_2) \Gg(t_2,t_1)]^{q/2-1}\Gg(t_1,t_2).
	\end{align}
	Over this quench, the Green's functions must be continuous; this holds also along the blue edge in Fig.~\ref{fig:protocol}(b): $g_{00}(t_1,0^-) = g_{01}(t_1,0^+)$ and $g_{01}(0^-,t_2) = g_{11}(0^+,t_2)$. As shown in \cite{Louw2022Feb,Eberlein:2017jb}, the large-$q$ dynamics is described by
	\begin{equation}
		\partial_{t_1} g(t_1,t_2)
		=\imath \epsilon(t_1) + 2\nint[t_1][t_2]{t_3} \vec{\Jj}(t_1)\cdot \vec{\Jj}(t_3) e^{g(t_1,t_3)}. \label{KB}
	\end{equation}
	Here the first derivative encodes the energy
	\begin{equation}
		\epsilon(t) = 2\Im \nint[-\infty][t]{t_3} \vec{\Jj}(t)\cdot \vec{\Jj}(t_3) e^{g(t,t_3)} \label{energyInt}
	\end{equation}
	which is conserved in each block. The proof that the above expression $\epsilon(t)$ is proportional to the energy density $\ex{H/N}$ follows from the generalized Galitskii-Migdal relation \cite{Louw2022Feb,Stefanucci2013Mar}. If we only have terms of order $q$ in $H(t)$ throughout the time evolution, then it is in fact proportional to the total energy, which is piecewise constant over the quench. Since this is the case under consideration, we have that $\epsilon$ is piecewise constant $\epsilon(t) = \epsilon_{0} \Theta(-t) + \epsilon_{1} \Theta(t)$. Knowing this, a second derivative of \eqref{KB} yields the Liouville-type equation
	\begin{equation}
		\p_{t_1} \p_{t_2} g(t_1,t_2) = 2 \vec{\Jj}(t_1)\cdot \vec{\Jj}(t_2) e^{g(t_1,t_2)}.\label{liouville}
	\end{equation}
	
	Since the system thermalizes instantaneously in the large $q$ limit \cite{Louw2022Feb,Eberlein:2017jb}, the Green's functions in the diagonal blocks ($b \in \{0,1\}$) depend solely on time differences
	\begin{equation}
		g_{b}(t_1-t_2) \equiv g_{bb}(t_1,t_2).
	\end{equation}
    In this notation the boundary conditions in Fig.~\ref{fig:protocol}(b) are
	\begin{equation}
		g_{01}(t,0) = g_{0}(t)\quad	g_{01}(0,t) = g_{1}(-t).
		\label{BCs}
	\end{equation}
	Further the Liouville equation \eqref{liouville} now becomes
	\begin{equation}
		\ddot{g}_{b}(t) = -2\Jj_{b}^2 e^{g_{b}(t)}. \label{diagonalLiouville}
	\end{equation}
	There are two known piecewise solutions to \eqref{liouville}, namely for the $I_{b} \times I_{b}$ blocks in Fig.~\ref{fig:protocol}(b)
	\begin{equation}
		e^{g_{b}(t)/2} = \frac{\lambda_{L,b}/2}{\Jj_{b} \cos(\pi v_{b}/2- \i \lambda_{L,b} t/2)}. \label{expg}
	\end{equation}
    Here $\lambda_{L,b}$ is the Lyapunov exponent of the system \cite{Maldacena2016Aug,Maldacena2016Nov,vanmanen}. The boundary condition $g(0)=0$ yields a closure relation
	\begin{equation}
		\lambda_{L,b} \equiv 2 \Jj_{b} \cos(\pi v_{b}/2). \label{sigma_s}
	\end{equation}
    Differentiating \eqref{expg} yields
	\begin{equation}
		\dot{g}_{b}(t) = -\i \lambda_{L,b} \tan([\pi v_{b} - \i\lambda_{L,b} t]/2), \label{gdot}
	\end{equation}
	a quantity that is related to the energy integral via \eqref{KB} $\dot{g}_{b}(0) = \imath \epsilon_{b}$, or explicitly
	\begin{equation}
		\epsilon_{b} = -2\Jj_{b} \sin(\pi v_{b}/2) = \Im \dot{g}_{b}(0).\label{epsilon_s}
	\end{equation}
	
	Substituting the second-order differential equation \eqref{liouville} into the energy integral \eqref{energyInt} yields the post-quench energy
	\begin{equation}
		\epsilon_{1} = \Im[\p_t  (g_{01}^*(0^-,t)-g_{01}^*(-\infty,t))+\dot{g}_{1}(0^+)-\dot{g}_{1}(t)].  \label{epsilon1first}
	\end{equation}
	From the boundary conditions \eqref{BCs} we know that $g_{1}(t)= g_{01}^*(0,t)$ and so these two terms cancel leaving
	\begin{equation}
		\epsilon_{1} = \Im[\dot{g}_{1}(0^+)-\p_t g_{01}^*(-\infty,t)]. \label{epsilon1}
	\end{equation}
	From \eqref{KB}, we note that $\imath\epsilon_{1} = \dot{g}_{1}(0^+)$; hence
	\begin{equation}
		\p_t \Im g_{01}(-\infty,t) = 0 \label{conditionInf}
	\end{equation}
	where the imaginary part vanishes identically for all $t$. This will turn out to be our key condition for relating the pre- and post-quench state variables. For instance we find the associated temperature before and after the quench by considering the KMS relation. This leads us to the imaginary time periodicity
	\begin{equation}
		\beta_{b} = \frac{2\pi v_{b}}{\lambda_{L,b}} = \frac{\pi v_{b}}{\Jj_{b} \cos(\pi v_{b}/2)}. \label{betai}
	\end{equation}	
    where $v \in [0,1]$. So $v=0$ corresponds to $T\equiv 1/\beta \to \infty$ and $v=1$ corresponds to $T=0$.
	Assuming some initial temperature $\beta_{0}$, there is only one unknown left, $\beta_{1}$, which is fully determined by $v_{1}$. As such what remains is to relate $v_{1}$ to $v_{0}, \Jj_{b}, c_{01}$.
	
	\subsection{Off-diagonal time block solution} \label{OffdiagSol}
	To obtain our thermal update rule, e.g., the relation between the initial and final temperatures, we require the off-diagonal solution in Fig.~\ref{fig:protocol}(b). Due to an SL$(2)$ symmetry in \eqref{liouville}, the solution may be parametrized as \cite{Eberlein:2017jb}
	\begin{equation}
		e^{g_{01}(t_1,t_2)} \equiv \frac{-\dot{h}_{0}(t_1)\dot{h}_{1}(t_2)}{c_{01} \Jj_{0} \Jj_{1} [h_{0}(t_1)-h_{1}(t_2)]^2}. \label{gb}
	\end{equation}
	The solution \eqref{gb} is invariant under the M\"obius transformations $h_{i} \to (a h_{i} + b)/(c h_{i} + d)$ with $ad-bc=1$, acting simultaneously on both $h_{0}$ and $h_{1}$. Therefore, without loss of generality, we may impose the three conditions
	\begin{equation}
		h_{0}(0) = 1, \qquad h_{1}(0) = 0, \qquad \dot{h}_{0}(0) = -1 .
		\label{gauge}
	\end{equation}
	Finally the boundary condition $e^{g_{01}(0,0)} = 1$ in \eqref{gb} fixes the last derivative $\dot{h}_{1}(0) = c_{01} \Jj_{0} \Jj_{1}$.
	
	Using \eqref{gdot} and \eqref{epsilon_s}, it will serve us to define
	\begin{equation} 
		V_{b}(t) = \frac{\dot{g}_{b}(0)-\dot{g}_{b}(t)}{2\Jj_{b}}
		= \frac{2 \Jj_{b}}{\lambda_{L,b}\coth(\lambda_{L,b}t/2) - \i\epsilon_{b}} ,
		\label{fdefnew}
	\end{equation}
    which satisfies $V_{b}(0)=0$ and, by \eqref{diagonalLiouville}, $\dot{V}_{b}(t) = \Jj_{b}e^{g_{b}(t)}$. 
	
	To solve for $h_{0}$ and $h_{1}$ we only require the boundary conditions \eqref{BCs}, namely that $e^{g_{01}(t,0)} =\dot{V}_{0}(t)/\Jj_0$ and $e^{g_{01}(0,t)} =\dot{V}_{1}(-t)/\Jj_1$. As such setting either $t_1=0$ or $t_2 = 0$ in \eqref{gb} together with \eqref{gauge} respectively yields
    \begin{align}
    \dot{V}_{1}(-t_2)/\Jj_1 &= \frac{\dot{h}_{1}(t_2)}{c_{01} \Jj_{0} \Jj_{1} [1-h_{1}(t_2)]^2} \notag\\
    &= \left(\frac{1}{c_{01} \Jj_{0} \Jj_{1} [1-h_{1}(t_2)]}\right)^{\prime},\\
    \dot{V}_{0}(t_1)/\Jj_0 &= -\dot{h}_{0}(t_1)/h_{0}(t_1)^{2} = [1/h_{0}(t_1)]^{\prime}.
    \end{align}
    Integrating the above yields the solutions
	\begin{equation}
	\frac{1}{1-h_{1}(t)} = 1 - c_{01} \Jj_{0} \,V_{1}(-t),\quad	\frac{1}{h_{0}(t)} = 1 + \frac{V_{0}(t)}{\Jj_{0}}. \label{hsolved}
	\end{equation}
	After some cancellations this leaves our central result
	\begin{align}
		e^{g_{01}(t_1,t_2)} =& \frac{e^{g_{0}(t_1)} e^{g_{1}(-t_2)}}{\left[1+c_{01} V_{0}(t_1)V_{1}(-t_2)\right]^2}.\label{gbnew}
	\end{align}
	
	\subsection{Closed quantum system heating} \label{secHeating}
    Our next goal is to use our newly derived off-diagonal solution \eqref{gbnew} to find the thermal update rule. We do this by focusing on the energy relation \eqref{epsilon_s} $\epsilon_{1} = \Im\dot{g}_{1}(0)$. Using \eqref{gbnew}, we find
	\begin{align*}
		\Im[\p_t g_{01}(-\infty,t)] =&  -\Im[\dot{g}_{1}(t)]\\&-2 \Im[\p_{t}\ln\left[1+c_{01} V_{0}(-\infty)V_{1}(-t)\right]].
	\end{align*}
	We next use condition \eqref{conditionInf}; this reduces to
    \begin{align*}
		\Im[\dot{g}_{1}(0)] =& -2 \Im[\p_{t}\ln\left[1+c_{01} V_{0}(-\infty)V_{1}(-t)\right]]\vert_{t=0}\\
		=& 2 \Im\left[\frac{c_{01} V_{0}(-\infty)\dot{V}_{1}(0)}{1+c_{01} V_{0}(-\infty)V_{1}(0)}\right]\\
		=&  2 c_{01}\Jj_1 \Im [ V_{0}(-\infty) ]
	\end{align*}
	where we have used $V_{1}(0) = 0$ and $\dot{V}_{1}(0) = \Jj_1$. Further from \eqref{fdefnew} $V_{b}(t\to-\infty) \to -e^{\imath \pi v_{b}/2}$ having made use of \eqref{sigma_s} and \eqref{epsilon_s}. This leaves $\epsilon_{1} = -2 c_{01} \Jj_{1} \sin(\pi v_{0}/2)$; hence
	\begin{equation}
		\sin(\pi v_{1}/2) = c_{01} \sin(\pi v_{0}/2), \quad c_{01} \equiv  \frac{\vec{\mathcal{J}}_0\cdot\vec{\mathcal{J}}_1}{\mathcal{J}_0 \mathcal{J}_1}. \label{vrel}
	\end{equation}
	Via \eqref{betai} using \eqref{vrel} we can then obtain the final temperature, $T_{1} = \Jj_{1} \cos(\pi v_{1}/2)/(\pi v_{1})$. The prefactor in \eqref{vrel} is nothing but the cosine of the angle $\vartheta$ between the two coupling vectors
	\begin{equation}
		c_{01} =  \frac{\vec{\mathcal{J}}_0\cdot\vec{\mathcal{J}}_1}{\mathcal{J}_0 \mathcal{J}_1} = \cos\vartheta \le 1,
		\label{cauchySchwarz}
	\end{equation}
	the inequality being Cauchy-Schwarz. Since $\sin(\pi v/2)$ is monotonically increasing on $v \in [0,1]$, it follows that $v_{1}\le v_{0}$, which leads to heating in every non-trivial case. The lack of cooling is to be expected from the second law of thermodynamics, given that we have a quench of a closed quantum system. The heating is directly set by how much the post-quench Hamiltonian is rotated away from the pre-quench one. In this sense the second law becomes a geometric statement, with the maximally non-commuting quench $\vartheta \to \pi/2$ heating the system to infinite temperature, $v_{1}\to 0$.
	
	The system is of course closed and thus the unitary evolution preserves von Neumann entropy and no full system heating can truly occur. The temperature here is thus defined locally (in flavor space) via observables as read off from the local two-point function. In this way the model is its own bath.
	
	\section{Conclusion and outlook} \label{secConclusion}
	
	In summary we have solved the large-$q$ SYK quench problem across all four time blocks for a quench between non-commuting $q/2$-body Hamiltonians. The non-equilibrium off-diagonal blocks in Fig.~\ref{fig:protocol}(b), previously accessible only numerically, are now given in closed form via \eqref{gbnew}. Energy conservation fixes the post-quench temperature, yielding the exact relation \eqref{vrel}. Further, a Cauchy-Schwarz inequality (Eq.~\eqref{cauchySchwarz}) bounds the prefactor by unity, and as such the system always heats by an amount controlled by the angle between the two coupling vectors.
	
	Our newly derived closed-form off-diagonal blocks now allow a direct comparison with finite-$q$ numerics \emph{during} the quench; the focus of the companion paper \cite{CompanionOsterkorn}. In said paper the non-thermal nature of these blocks is used to reconcile large-$q$ instantaneous thermalization with the finite relaxation rates of the effective temperature.
	
	A natural extension is a sequence of quenches and thus also the possibility for a continuum limit solution. Such solutions would allow one to analytically study phase transitions driven by continuous heating.
	
	A second extension is to study possible cooling via anti-Hermitian quenches. Throughout we assumed Hermiticity, reflected in the couplings $\vec{\Jj}$ being real, which stems from the relation $Z^{\bm{\nu}}_{\bm{\mu}} = (Z^{\bm{\mu}}_{\bm{\nu}})^{*}$. The opposite extreme gives an anti-Hermitian Hamiltonian. This is not necessarily ill-defined: the antiunitary particle-hole symmetry of our model is of $PT$ type~\cite{Kanazawa:2017be}, so
the spectrum of such a deformation is either real or organized into complex-conjugate pairs. It suggests an SYK open-system reading \cite{Prosen2008Apr,Sa2022Jun,Kulkarni2022Aug,Kawabata2023Aug,Zanoci2022Apr,Cheipesh2021Sep}, since non-Hermitian evolution arises as the no-jump sector of a Lindblad equation. This raises the question of which open system it corresponds to, i.e., which bath. We can already exclude a Markovian bath with a unique stationary state, since the heating and cooling we observe are dependent on the initial temperature. We leave this to future work.
	\begin{acknowledgments}
		Much of this work was carried out during a research stay of J.C.L. at the California Institute of Technology, whose hospitality is gratefully acknowledged. J.C.L.\ thanks Iliya Esin for discussions at an early stage of this work. During this time J.C.L. was employed as a postdoc at the University of G\"ottingen. This work was in part funded by the Deutsche Forschungsgemeinschaft (DFG, German Research Foundation) --- Project No. 217133147/SFB 1073, Project B03. J.C.L. would also like to thank Johan du Buisson and Jonas Loy for helpful feedback on the calculations done in this work. 
	\end{acknowledgments}

	\bibliography{references}

\appendix
\section{Green's functions relations}
\label{AppGreen}

    Our object of focus throughout this work is the flavor-averaged correlation functions ordered along the Keldysh contour $\Cc$
	\begin{equation}
		\Gg(t_1,t_2) \equiv \frac{-\imath}{N} \sum_{\alpha=1}^{N}\ex{\Tt_{\Cc} \hat{c}_{\alpha}(t_1) \hat{c}_{\alpha}^{\dag}(t_2)}. \label{AppGdef}
	\end{equation}
    Note that in \eqref{AppGdef} we are using the standard real time Green's function convention above. This is related to the Euclidean convention via $\Gg = \imath \Gg_{E}$, $\Upsigma \equiv \Sigma_E/\imath$, which is typically reserved for imaginary time studies. Here $\Tt_\Cc$ is the time ordering operator w.r.t. $\Cc$. Over the Keldysh contour, the correlations may be separated piece-wise into the greater- and lesser-than versions
	\begin{equation}
		\Gg(t_1,t_2) = \Theta_\Cc(t_1,t_2)\Gg^>(t_1,t_2) + \Theta_\Cc(t_2,t_1) \Gg^<(t_1,t_2).
	\end{equation}
	In the large $q$ limit they are captured by the expressions
	\begin{equation}
		\Gg^{\gtrless}(t_1,t_2)  = \mp \frac{\imath}{2} e^{ g^\gtrless(t_1,t_2)/q}, \label{largeqformApp}
	\end{equation}
	meaning that the interactions enter in at first order in $1/q$. The structure \eqref{largeqformApp} is chosen to yield the equal time condition $g^{\gtrless}(t,t)=0$. We focus on the symmetric function
	\begin{align}
		g(t_1,t_2) &\equiv \frac{g^>(t_1,t_2) + g^<(t_2,t_1)}{2} \label{gB}
	\end{align}
    Although we are considering the charged/complex SYK case \cite{Sachdev2015Nov,Davison2017Apr,Gu2020Feb}, we will focus on half filling, which overlaps with the Majorana SYK case in many ways. Like the Majorana case, we have $g^>(t_1,t_2) = g^<(t_2,t_1)$, hence $g(t_2,t_1)^* = g(t_1,t_2)$ in our charge-neutral case \cite{Eberlein:2017jb,Louw2022Feb}.
	
\end{document}